\documentclass[12pt]{article}
\usepackage{style-Ariel}

\title{Evaluating Pest Management Strategies:\\
A Robust Method and Its Application to Strawberry Disease Management\thanks{Selected paper prepared for presentation at the 2019 Agricultural \& Applied Economics Association Annual Meeting, Atlanta, GA, July 21-23.}}

\author{Ariel Soto-Caro\thanks{Copyright 2019 by Soto-Caro, Wu, Guan, and Peres. All rights reserved. Readers may make verbatim copies of this document for non-commercial purposes by any means, provided that this copyright notice appears on all such copies.}\\ \href{mailto:asotocaro@ufl.edu}{\texttt{asotocaro@ufl.edu}} 
\and Feng Wu\\ \href{mailto:fengwu@ufl.edu}{\texttt{fengwu@ufl.edu}} 
\and Zhengfei Guan\\ \href{mailto:guanz@ufl.edu}{\texttt{guanz@ufl.edu}} 
\and Natalia Peres\\ \href{mailto:nperes@ufl.edu}{\texttt{nperes@ufl.edu}}}

\date{August 7, 2019}

\begin{document}
{\setstretch{.8}

\maketitle
\begin{abstract}
    Farmers use pesticides to reduce yield losses. The efficacies of pesticide treatments are often evaluated by analyzing the average treatment effects and risks. The stochastic efficiency with respect to a function is often employed in such evaluations through ranking the certainty equivalents of each treatment. The main challenge of using this method is gathering an adequate number of observations to produce results with statistical power. However, in many cases, only a limited number of trials are replicated in field experiments, leaving an inadequate number of observations. In addition, this method focuses only on farmer’s profit without incorporating the impact of disease pressure on yield and profit. The objective of our study is to propose a methodology to address the issue of an insufficient number of observations using simulations and take into account the effect of disease pressure on yield through a quantile regression model. We apply this method to the case of strawberry disease management in Florida.
\end{abstract}
}
\section{Introduction}
Farmers use pesticides to reduce yield losses. The efficacies of pesticide treatments are often evaluated by analyzing the average treatment effects and risks (i.e., the distribution of treatment effects) based on experimental trails with limited replicates. The stochastic efficiency with respect to a function (SERF) method as presented by Hardaker et al. (\citeyear{Hardaker2004}) is currently widely used in this field because SERF takes into account the decision-maker’s risk profile. SERF methodology is used to rank treatments in either controlled or non-controlled experiments. Using SERF requires only a small set of data and makes minimum assumptions. Researchers need define a utility function of wealth or profit that reflects the farmer’s risk preference. After the key data such as production yield, cost, and profit for each treatment under evaluation are obtained, researchers can compute the certainty equivalent (CE) and the risk premium (RP). The CE measure represents the sure amount that generates the same utility as with a risky outcome \citep{Hardaker2004}. As a result, SERF provides farmers with a very intuitive tool to make risk-efficient decisions by simply choosing the best-ranked treatment based on the CEs.

However, the main challenge of this method is to gather an adequate number of observations in order to produce results with statistical power. Because only a limited number of trials are replicated in field experiments and thus producing a limited number of observations, the results are questionable due to the lack of statistical power. In addition, this method ignores the impact of disease pressure/weather factors on yield and profit. What happens if ignoring the influence of these exogenous variables on the risk-efficiency of the treatment? Is the CE ranking statistically robust? To answer these questions, we propose to estimate the relationship between production yield and exogenous variables that impact farming outcome to test the effect of these variables on the CE ranking under simulation.

To scrutinize the CE approach, we analyze the case of \textit{Botrytis cinerea Pers}. (BCP) in Florida strawberries. Florida is the second largest strawberry producing state in the United States, having a farm gate value of approximately \$282 million in 2018 \citep{USDA-NASS2019}. Florida strawberry growers face many challenges caused by Botrytis. In the past decade, Florida strawberry yield declined significantly by 36\%, from 320 cwt/acre in 2007 to 205 cwt/acre in 2016; some reports have estimated even higher yield losses (50–70\%) (Cordoba et al., \citeyear{Cordova2014}; Legard et al., \citeyear{Legard2003}). Botrytis causes significant losses of pre- and post-harvest fruit because the disease can develop both in the field and during storage and transportation. Under a variety of unfavorable environmental conditions, it could become one of the most challenging pathogens to control \citep{Braun1987,MacKenzie2012a,MacKenzie2012b}.

Botrytis is a pest that affects more than 250 different crops and is highly resistant to pesticides (Gobeil-Richard et al., \citeyear{Gobeil-Richard2016}). Because many treatments lose their effectiveness over time, research in pest disease management and pesticides development becomes even more essential for the future of agriculture \citep{Frisvold2019}.

In this paper, we propose an alternative approach to SERF to compute CE under different scenarios. This will allow the analyst to understand the effect of disease pressure on the CE measure under different yield levels. We present this approach in an application to the case of the Florida strawberry disease management.

\section{Materials and Methods}

In this section, we discuss the dataset used to describe the procedure for computing the standard CE measure, and explain our alternative methodology based on simulations.

\subsection{Data}
The computation of the regular CE only requires some basic elements. For the simulation, we employed the Botrytis Incidence Index (BII) to measure the disease pressure (Botrytis).

Yield data were collected from strawberry field trials on a commercial farm in Plant City, Florida. In these trials, three fungicide treatments (Fracture, Milstop, and Serenade) were tested to protect strawberries from botrytis. Each treatment had four replications in each season, generating in total eight profit observations per treatment in two seasons. Strawberries were harvested 24 times over one season. 

The spraying costs varied across the treatments (Table \ref{tab:1}). The remaining cost items (e.g., overhead, harvest, and marketing costs) were obtained from the cost budget by Guan et al. (\citeyear{Guan2017}) for season 2012-13. Using the Productivity Price Index, we updated these costs for seasons 2014-15 and 2015-16 \citep{USBLS2019}. Price data were obtained from the Agricultural Market Service, the U.S. Department of Agriculture \citep{USDA2019}. 

\begin{figure}[H]
\centering
    \includegraphics[scale=.75]{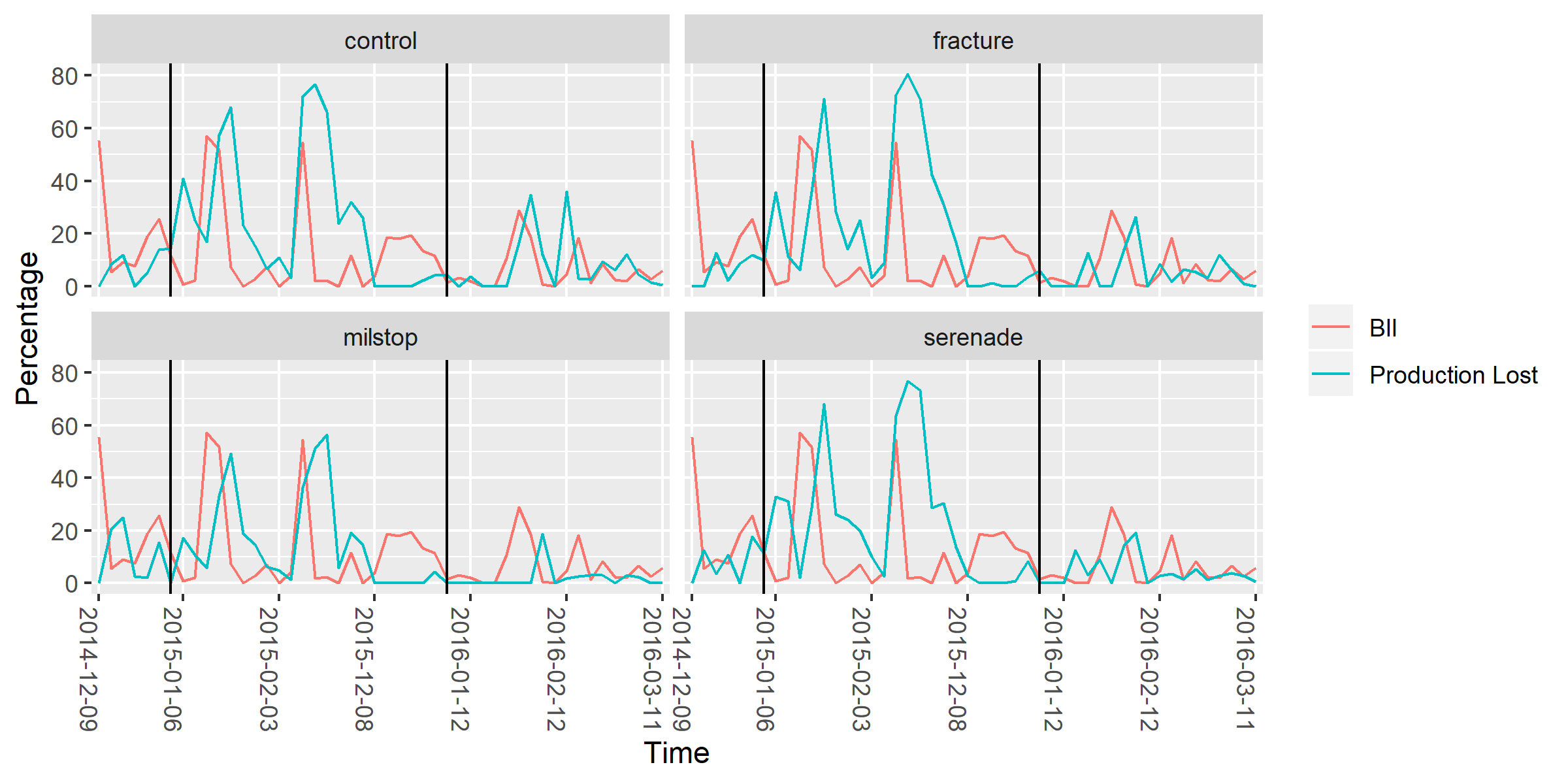}
\caption{Daily behavior of the Botrytis Infection Index and production loss of strawberry in trial fields.}
\label{fig:losses}
\end{figure}

The botrytis infection index was obtained from AgroClimate\footnote{http://agroclimate.org/tools/sas}, a scientific group of the Institute of Food and Agricultural Sciences at the University of Florida that develops agroclimatology tools to keep farmers informed. The dynamics of the BII over two seasons is presented in Figure \ref{fig:losses}. The index is higher for the 2014-15 season than for the 2015-16 season. Also,  there is a high correlation between BII and the percentage of crop losses, indicating this disease could be a determinant of strawberry productivity in Florida. 

\subsection{Certainty equivalent}

For a risk-averse farmer, the estimated CE will be less than the expected money value (EMV), and its difference (EMV-CE) is the risk premium. The ordering of risky alternatives by CE is the same as ordering them by utility values (Hardaker et al., \citeyear{Hardaker2004}). Assuming a power utility function (PU), the most widely used functional form in the empirical analysis and recommended for multiple-year analysis \citep{Richardson2008}, the CE is
\begin{equation}
    CE = \begin{cases}
    \overline{\pi} & \text{if } RAC=0\\
    \left[ \frac{\sum^R_{r=1}\left(\pi^R_{n,M}+w_0\right)^{RAC}}{R}\right]^{\frac{1}{RAC}}-w_0 & \text{otherwise}
    \end{cases}
    \label{eq:1}
\end{equation}
where $\pi^R$ is the profit for the replicate $R$, $RAC$ is the relative Risk Aversion Coefficient, and $w_o$ is initial wealth. The $\pi$ is the sum of the revenue (price $P$ times the yield $Y$ for each treatment $M$) at every harvest $n$, minus the total cost ($TC$).
\begin{equation}
    \pi^R_{n,M} = \sum_{n=1}^N \left(P_n \cdot Y_{n,M}^R - TC \right)
    \label{eq:2}
\end{equation}

The range of $RAC$ is set from $0.5$ to $4$, representing `hardly risk averse at all' to `extreme risk aversion' \citep{Anderson1992}. 

\subsection{Alternative procedure}

Disease pressure has a direct effect on farmers’ profit, while weather has a direct effect on BII. We introduced the exogenous BII index into a quantile-regression model to estimate the relationship between weather factors and strawberry yield, which was used to simulate sufficient yield observations under different weather scenarios. Then these simulated yield estimates rather than the actual observations were used to compute CE in order to increase the statistical power. The quantile regression is considered a mixture of parametric and non-parametric estimation. It is parametric because it has a functional form and is non-parametric in the sense that it allows the parameters to vary across quantiles \citep{Chavas2015}. The quantile regression has two advantages: (1) we can analyze different levels of yield distributions (one for each quantile) and (2) we can use simulated values of the exogenous variables to check the treatment’s sensitivity to the disease.

The following estimation was run for a quantile regression:
\begin{equation}
    Yield_t = \beta_0 + \beta_1 Yield_{t-1} + \beta_2 BII_t + \beta_3 t +  \sum_{i=1}^M \gamma_i D_i + t \sum_{i=1}^M \delta_i D_i
    \label{eq:3}
\end{equation}
$D$ is a dummy variable for each treatment $M=\{\text{Fracture,Milstop,Serenade}\}$. With the estimated model, we generated 100 yield simulations for three different BII levels: low levels of incidence disease (10–30\%), medium level of incidence (40–60\%), and high level of incidence (70–90\%).

The simulated yield was substituted into equation (\ref{eq:2}) to generated profit for the CE calculation,  

\begin{equation}
    \widetilde{\pi}_{n,M} = \sum_{n=1}^N \left( P_n \cdot Y_{n,M}(\widetilde{BII})-TC_{n,M} \right)
    \label{eq:4}
\end{equation}
where $\widetilde{BII}$ is the BII estimate. The simulation also considered three additional situations, where the historical-yield losses were low due to the BCP, meaning that we use a low yield quantile ($0.2$), a mid (average) yield quantile ($0.5$), and a high yield quantile ($0.8$). As a result, we performed a total of nine scenarios of combining the low/mid/high levels of yield and low/mid/high levels of botrytis incidence.

\section{Results}

\begin{figure}[H]
\centering
\includegraphics[scale=.55]{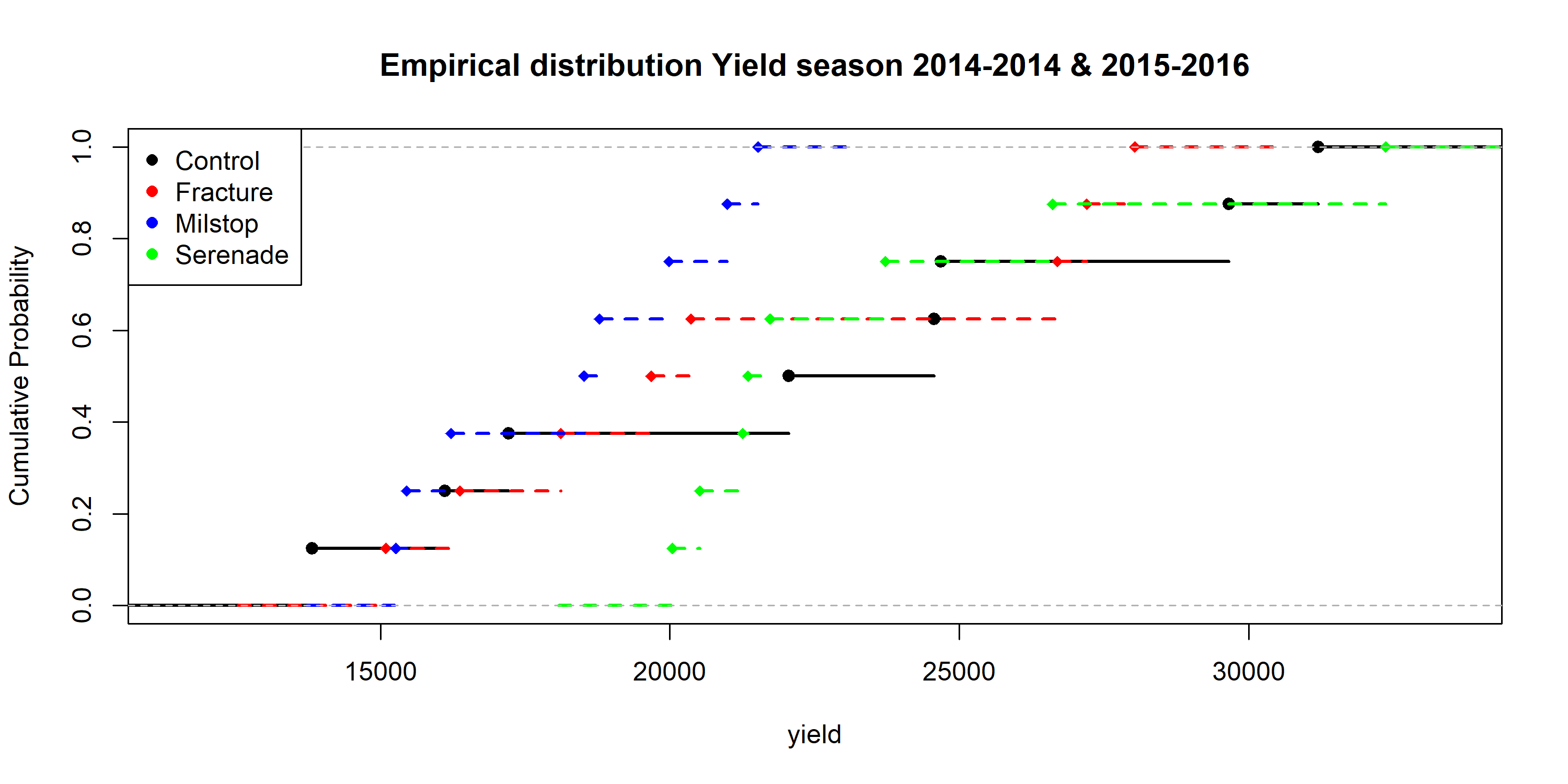}
\caption{Empirical distribution of yield for each treatment, seasons 2014-15 and 2015-16.}
\label{fig:empdist}
\end{figure}

The traditional stochastic dominance method does not show a clear difference of the yield distribution among the treatments (Figure \ref{fig:empdist}); given there is no clear risk-efficiency dominance, the implementation of CE is necessary to help decide which treatment is optimal.

\subsection{Regular CE computation}

\begin{figure}[H]
    \centering
    \begin{subfigure}{0.5\linewidth}
        \centering
        \includegraphics[scale=.5]{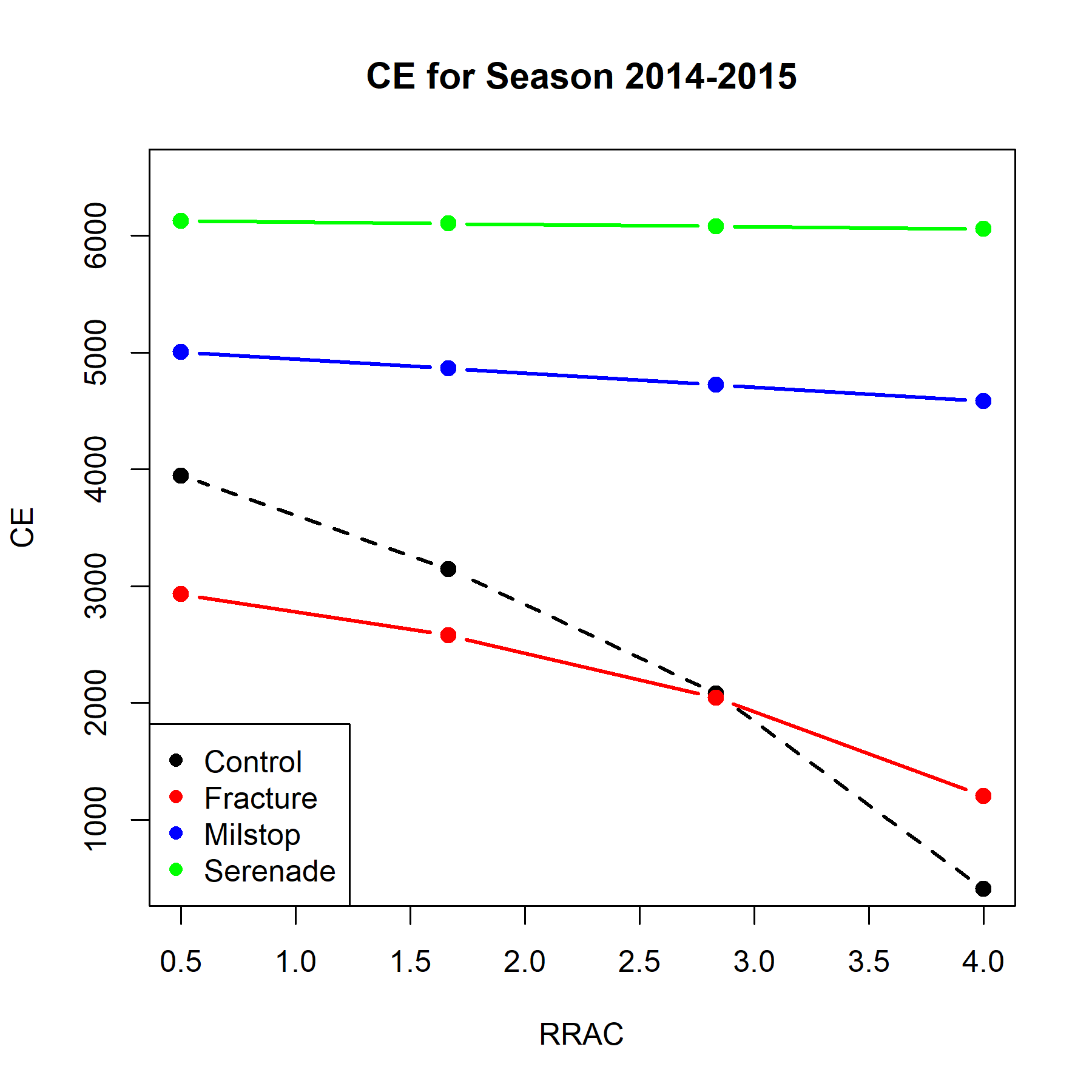}
        \caption{Season 2014-2015.}
        \label{fig:31}
    \end{subfigure}%
    ~ 
    \begin{subfigure}{0.5\linewidth}
        \centering
        \includegraphics[scale=.5]{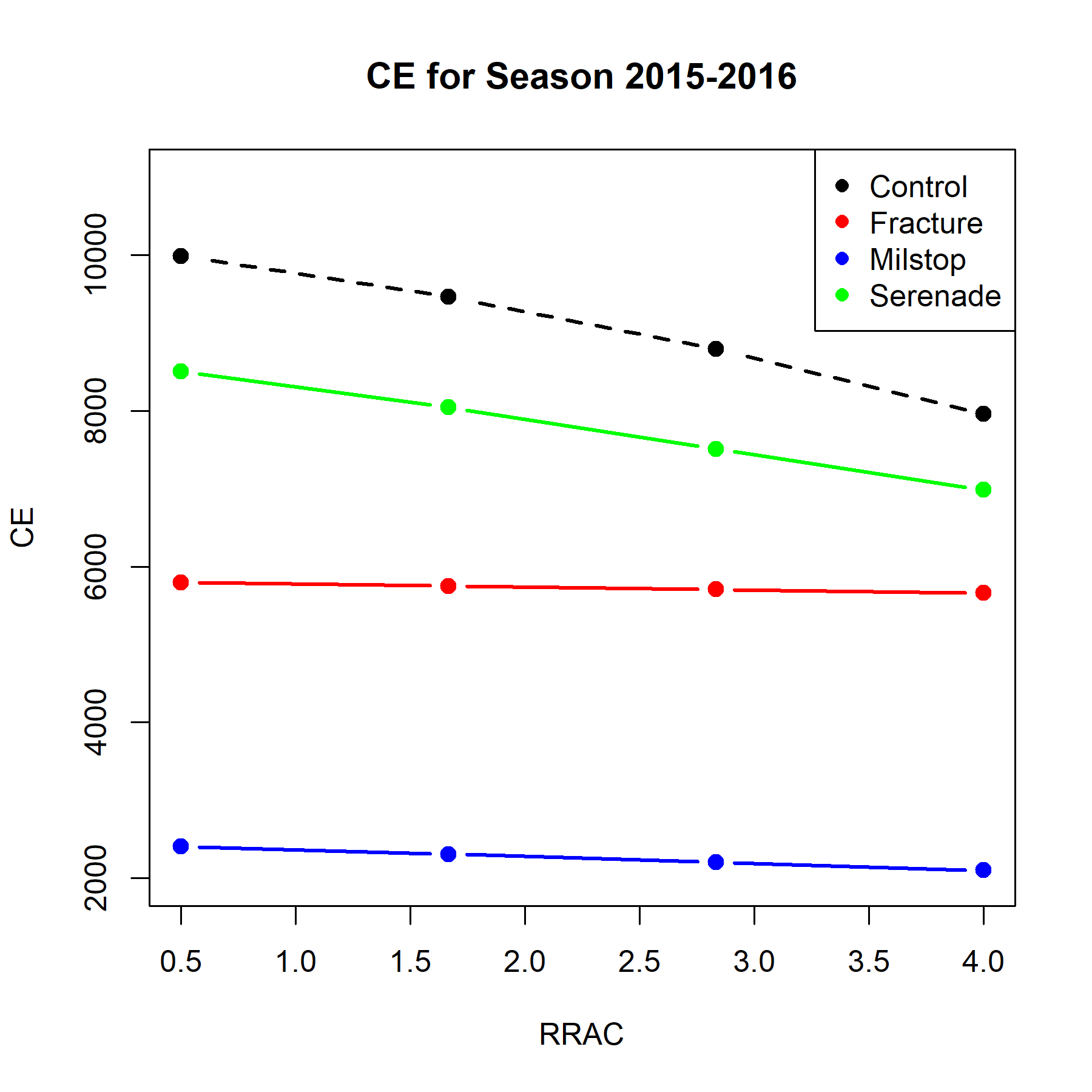}
        \caption{Season 2015-2016.}
        \label{fig:32}
    \end{subfigure}
    \\
    \begin{subfigure}{\linewidth}
        \centering
        \includegraphics[scale=.62]{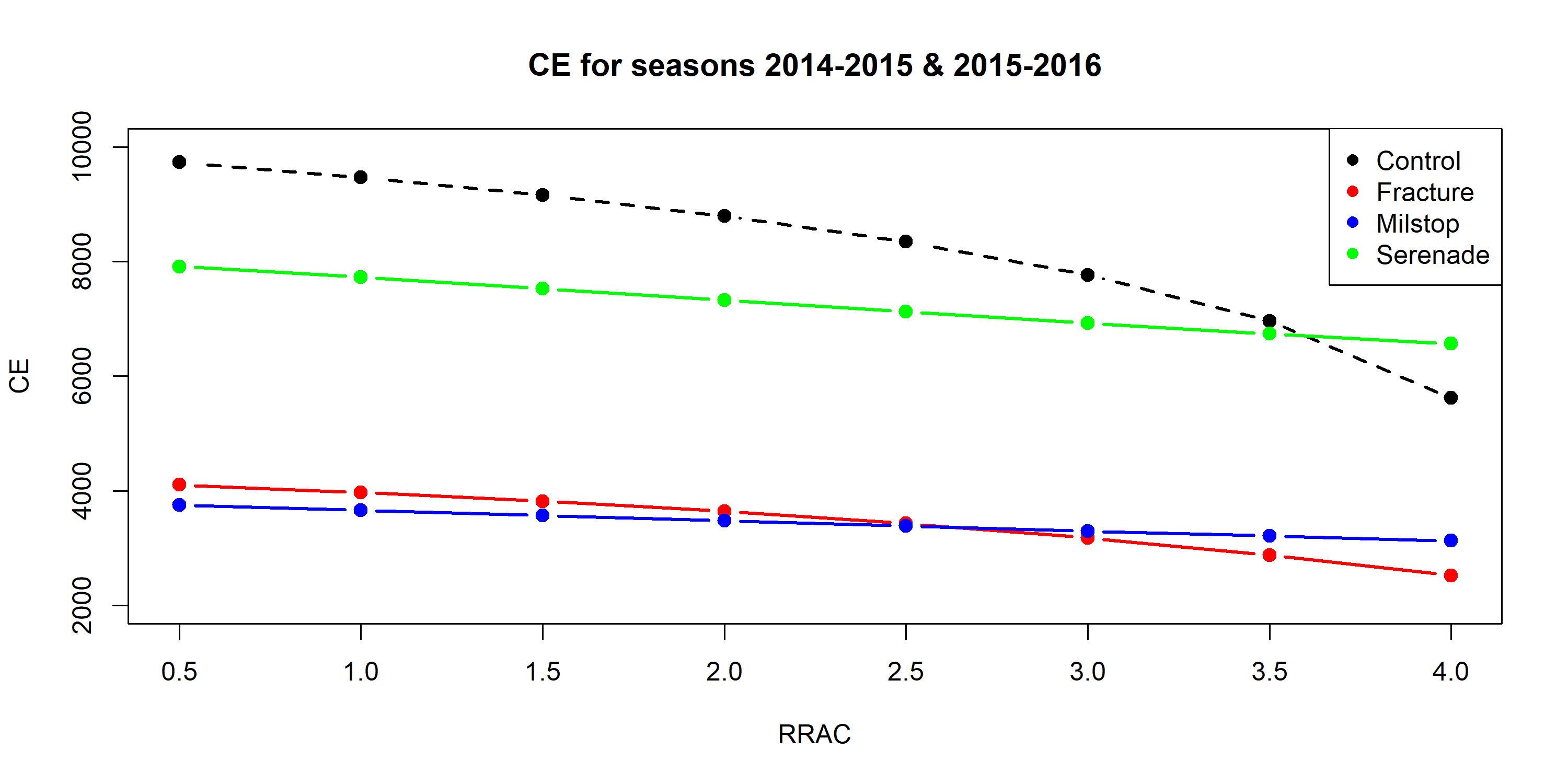}
        \caption{CE for the two seasons jointly.}
        \label{fig:33}
    \end{subfigure}
    \caption{Certainty Equivalents using current data.}
    \label{fig:3}
\end{figure}

The profit is computed for each treatment in each season (Table \ref{tab:1}) and then used to calculate CE (Figure \ref{fig:3}). CE ranking provides highly different recommendations: Serenade is the most risk-efficient for season 2014-15, and no treatment is better than the control case for season 2015-16. Considering two seasons together, Serenade is more efficient than the control case when the farmer is more risk-averse, or when the relative RAC is over $3.5$.

\subsection{Alternative  procedure}

\begin{figure}[H]
\centering
\includegraphics[scale=.4]{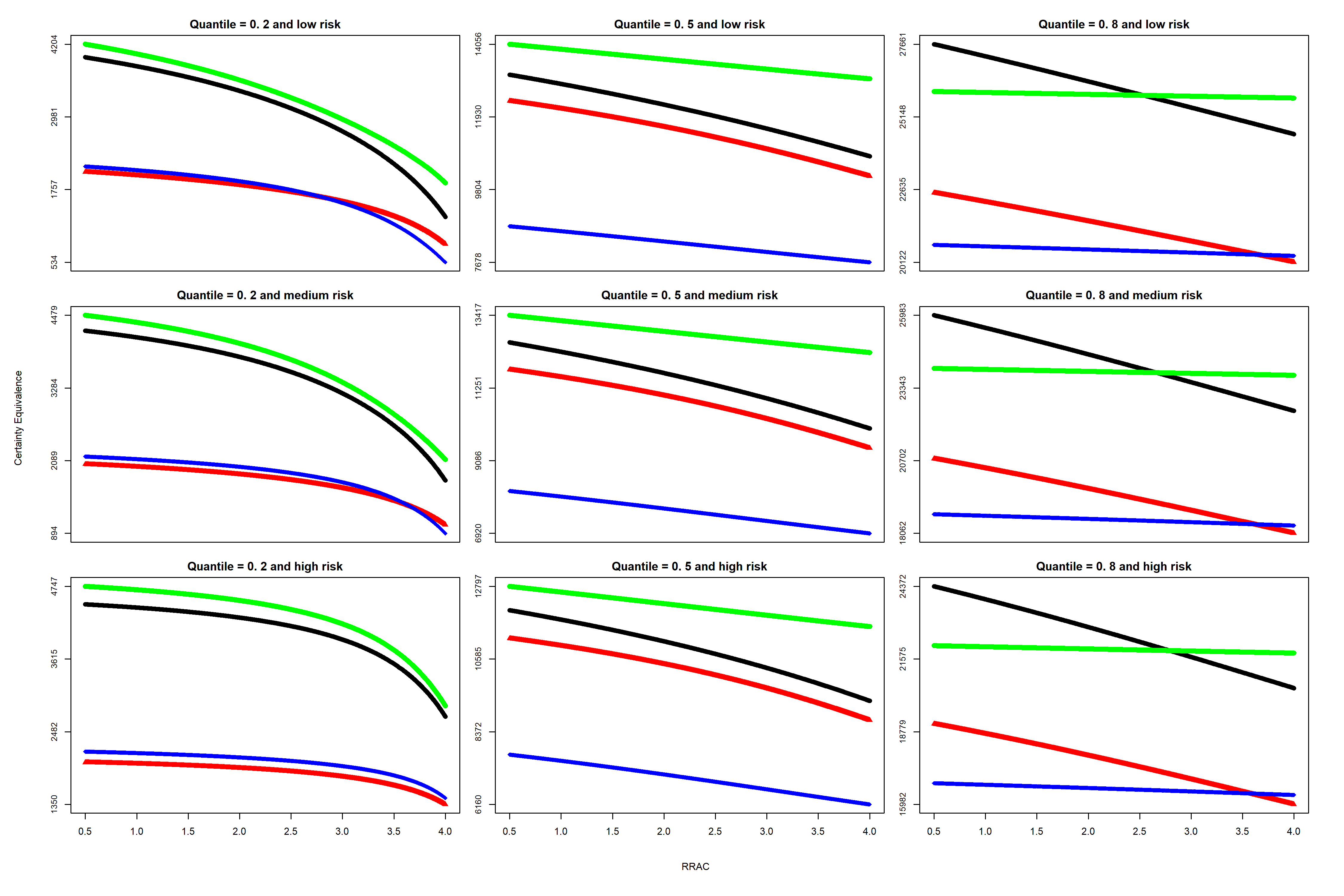}
\caption{Simulated CE (color lines are the same as previous figures).}
\label{fig:sim}
\end{figure}

Equation (\ref{eq:3}) is estimated with 768 observations (48 harvests $\times$ 4 replicates $\times$ 4 treatments); its coefficients are presented in Table \ref{tab:2}. With the parameters for each quantile, the predicted yield are simulated assuming specific levels of BII. This procedure is simulated 100 times for nine scenarios involving the combination of three disease incidence levels (low, medium, and high) with three levels of yield (low, medium and high). The CEs for each scenario are presented in Figure \ref{fig:sim}, where each row represents a different risk level based on the Botrytis incidence of low-to-high levels from top to bottom, and each column represents a different level of yield of low-to-high from left to right. The simulation results show that how the change in the disease incidence affects the level of profitability of each treatment, and how the quantile level of the yield affects the shape and slope of the CE directly and the ranking among treatments. This is consistent with the results presented in Figure \ref{fig:3} for most scenarios. In some scenarios, however, recommendations are different depending on the level of risk aversion. 

\section{Discussion and Conclusions}

In this study, we address the concern of statistical power by computing CE using simulated yield estimates rather than the actual observations. We use an estimated quantile regression to generate sufficient predicted yield for the CE approach. Further, the regression model incorporated the effect of disease incidents on yield. This approach has two advantages: (1) we can analyze the varying effects of disease on yield at different locations in the yield distribution, and (2) we can examine the treatment's sensitivity to disease pressure. The CE computed by our method can be used to present more clear results between different weather conditions or disease pressures, and between different levels of yield losses. Hence, it is a more reliable tool to find risk-efficiency treatments under different circumstances.

Unlike the basic CE recommendation, we found that Serenade is always risk-efficient in the lower quantiles; however, Serenade is chosen only in the higher quantiles when decision-makers are more risk-averse. At all risk averse levels, the control case is better than the Fracture or Milstop treatment. A sensitivity analysis computing the CEs with lower and higher prices show the same conclusion (see the Appendix).

\medskip
\bibliography{references.bib} 

\newpage

\begin{table}[H]
  \centering
  \caption{Descriptive statistics of the main variables.}
  \label{tab:1}
  \scalebox{.8}{
    \begin{tabular}{rlrrrrrr}
    \hline
    \multicolumn{1}{c}{\textbf{Treatment}} & \multicolumn{1}{c}{\textbf{Statistic}} & \multicolumn{1}{c}{\textbf{Unit}} & \multicolumn{1}{c}{\textbf{Rep}} & \multicolumn{1}{c}{\textbf{Mean}} & \multicolumn{1}{c}{\textbf{St. Dev.}} & \multicolumn{1}{c}{\textbf{Min}} & \multicolumn{1}{c}{\textbf{Max}} \\ \hline \hline

    \multicolumn{1}{l}{Control} & Total Spraying Cost & \multicolumn{1}{c}{\$} & \multicolumn{1}{c}{8} & 0     & 0     & 0     & 0 \\

          & Total Yield & \multicolumn{1}{c}{lb} & \multicolumn{1}{c}{8} & 22,411.20 & 6,325.90 & 13,819 & 31,201 \\

          & Total Profit & \multicolumn{1}{c}{\$} & \multicolumn{1}{c}{8} & 5,869.60 & 4,609.90 & -464.1 & 12,744.10 \\

    \multicolumn{1}{l}{Fracture} & Total Spraying Cost & \multicolumn{1}{c}{\$} & \multicolumn{1}{c}{8} & 1,777.40 & 144.5 & 1,642.30 & 1,912.60 \\

          & Total Yield & \multicolumn{1}{c}{lb} & \multicolumn{1}{c}{8} & 21,444.80 & 5,146.90 & 15,096 & 28,032 \\

          & Total Profit & \multicolumn{1}{c}{\$} & \multicolumn{1}{c}{8} & 4,138.50 & 2,644.10 & 22.2  & 7,932.70 \\

    \multicolumn{1}{l}{Milstop} & Total Spraying Cost & \multicolumn{1}{c}{\$} & \multicolumn{1}{c}{8} & 2,346.80 & 190.8 & 2,168.30 & 2,525.20 \\

          & Total Yield & \multicolumn{1}{c}{lb} & \multicolumn{1}{c}{8} & 18,343.30 & 2,460.70 & 15,269.00 & 21,524.10 \\

          & Total Profit & \multicolumn{1}{c}{\$} & \multicolumn{1}{c}{8} & 3,699.20 & 2,549.80 & 1,299.10 & 8,709.80 \\

    \multicolumn{1}{l}{Serenade} & Total Spraying Cost & \multicolumn{1}{c}{\$} & \multicolumn{1}{c}{8} & 2,288.80 & 186.1 & 2,114.80 & 2,462.90 \\

          & Total Yield & \multicolumn{1}{c}{lb} & \multicolumn{1}{c}{8} & 23,450.40 & 4,172.50 & 20,045.00 & 32,363.00 \\

          & Total Profit & \multicolumn{1}{c}{\$} & \multicolumn{1}{c}{8} & 6,619.80 & 1,918.40 & 4,479.70 & 10,633.90 \\
\hline
          & Price s. 2014-2015 &       &       & 14    & 6.61  & 6.9   & 27.9 \\

          & Price s. 2015-2016 &       &       & 22.8  & 7.73  & 10.2  & 31.4 \\
    \hline
    \end{tabular}}
\end{table}%

\begin{table}[H]
  \centering
  \caption{Coefficients estimated for the panel quantile regression.}
  \label{tab:2}
  \scalebox{.8}{
    \begin{tabular}{rrrrrrrrrr}\hline
          & \multicolumn{1}{l}{\textbf{Q = 0.1}} & \multicolumn{1}{l}{\textbf{Q = 0.2}} & \multicolumn{1}{l}{\textbf{Q = 0.3}} & \multicolumn{1}{l}{\textbf{Q = 0.4}} & \multicolumn{1}{l}{\textbf{Q = 0.5}} & \multicolumn{1}{l}{\textbf{Q = 0.6}} & \multicolumn{1}{l}{\textbf{Q = 0.7}} & \multicolumn{1}{l}{\textbf{Q = 0.8}} & \multicolumn{1}{l}{\textbf{Q = 0.9}} \\ \hline \hline
    \multicolumn{1}{l}{Constant} & -73.17 & -38.99 & 28.33 & 97.46 & \multicolumn{1}{l}{110.25**} & \multicolumn{1}{l}{112.88*} & \multicolumn{1}{l}{199.32**} & \multicolumn{1}{l}{228.73**} & \multicolumn{1}{l}{370.58**} \\
          & -58.92 & -34.19 & -71.45 & -67.72 & -47.22 & -62.09 & -85.31 & -115.45 & -151.18 \\
    \multicolumn{1}{l}{Yield (t-1)} & \multicolumn{1}{l}{0.4***} & \multicolumn{1}{l}{0.5***} & \multicolumn{1}{l}{0.62***} & \multicolumn{1}{l}{0.67***} & \multicolumn{1}{l}{0.77***} & \multicolumn{1}{l}{0.85***} & \multicolumn{1}{l}{0.87***} & \multicolumn{1}{l}{0.87***} & \multicolumn{1}{l}{0.92***} \\
          & -0.03 & -39.15 & -79.1 & -85.35 & -62.28 & -71.93 & -91.29 & -114.61 & -206.38 \\
    \multicolumn{1}{l}{BII} & -29.44 & -20.77 & -166.09 & -207.25 & -69.09 & -137.29 & -190.02 & -185.92 & -265.35 \\
          & -51.45 & -53.05 & -81.02 & -75.66 & -45.33 & -84.35 & -111.88 & -137.53 & -162.49 \\
    \multicolumn{1}{l}{t} & 0.06  & 0.36  & 0.25  & 0.2   & 0.17  & 0.6   & 0.84  & \multicolumn{1}{l}{3.05*} & \multicolumn{1}{l}{6.08***} \\
          & -0.56 & -45.36 & -84.72 & -102.95 & -105.04 & -129.29 & -133.48 & -113.61 & -151.61 \\
    \multicolumn{1}{l}{Fracture} & 20.36 & \multicolumn{1}{l}{83.1**} & 10.93 & -11.83 & -4.76 & 13.86 & -24.63 & -32.06 & -105.32 \\
          & -75.21 & -39.15 & -79.1 & -85.35 & -62.28 & -71.93 & -91.29 & -114.61 & -206.38 \\
    \multicolumn{1}{l}{Milstop} & \multicolumn{1}{l}{121.05**} & \multicolumn{1}{l}{93.38*} & 75.34 & 42.07 & 14.09 & 14.16 & 59.62 & 102.4 & 7.15 \\
          & -58.82 & -53.05 & -81.02 & -75.66 & -45.33 & -84.35 & -111.88 & -137.53 & -162.49 \\
    \multicolumn{1}{l}{Serenade} & 48.92 & 34.33 & -10.45 & 1.07  & 33.17 & 137.89 & \multicolumn{1}{l}{276.63**} & \multicolumn{1}{l}{316.23***} & \multicolumn{1}{l}{275.71*} \\
          & -60.55 & -45.36 & -84.72 & -102.95 & -105.04 & -129.29 & -133.48 & -113.61 & -151.61 \\
    \multicolumn{1}{l}{t x Serenade} & -0.25 & -0.05 & 0.28  & -0.05 & 0.02  & -1.03 & -2.02 & \multicolumn{1}{l}{-3.94**} & \multicolumn{1}{l}{-5.55***} \\
          & -0.74 & -0.68 & -0.82 & -0.92 & -0.93 & -1.15 & -1.41 & -1.79 & -1.31 \\
    \multicolumn{1}{l}{t x Fracture} & 0.17  & -0.65 & 0.12  & 0.41  & 0.31  & 0.09  & 0.6   & -0.69 & -1.51 \\
          & -0.64 & -0.61 & -0.9  & -0.82 & -0.67 & -0.94 & -1.33 & -1.96 & -1.76 \\
    \multicolumn{1}{l}{t x Milstop} & -0.62 & -0.87 & -0.74 & -0.63 & -0.3  & -0.28 & -0.92 & -1.97 & \multicolumn{1}{l}{-2.82*} \\
          & -0.57 & -0.64 & -0.78 & -0.72 & -0.6  & -0.91 & -1.41 & -2.01 & -1.64 \\ \hline
    \end{tabular}}
\end{table}%
\noindent
\footnotesize{\textbf{Note:} Standard errors in parenthesis. *, **, and *** represent significance level to 0.1, 0.05, and 0.01 respectively.}


\newpage


\section*{Appendix}

\begin{figure}[H]
\includegraphics[scale=.35]{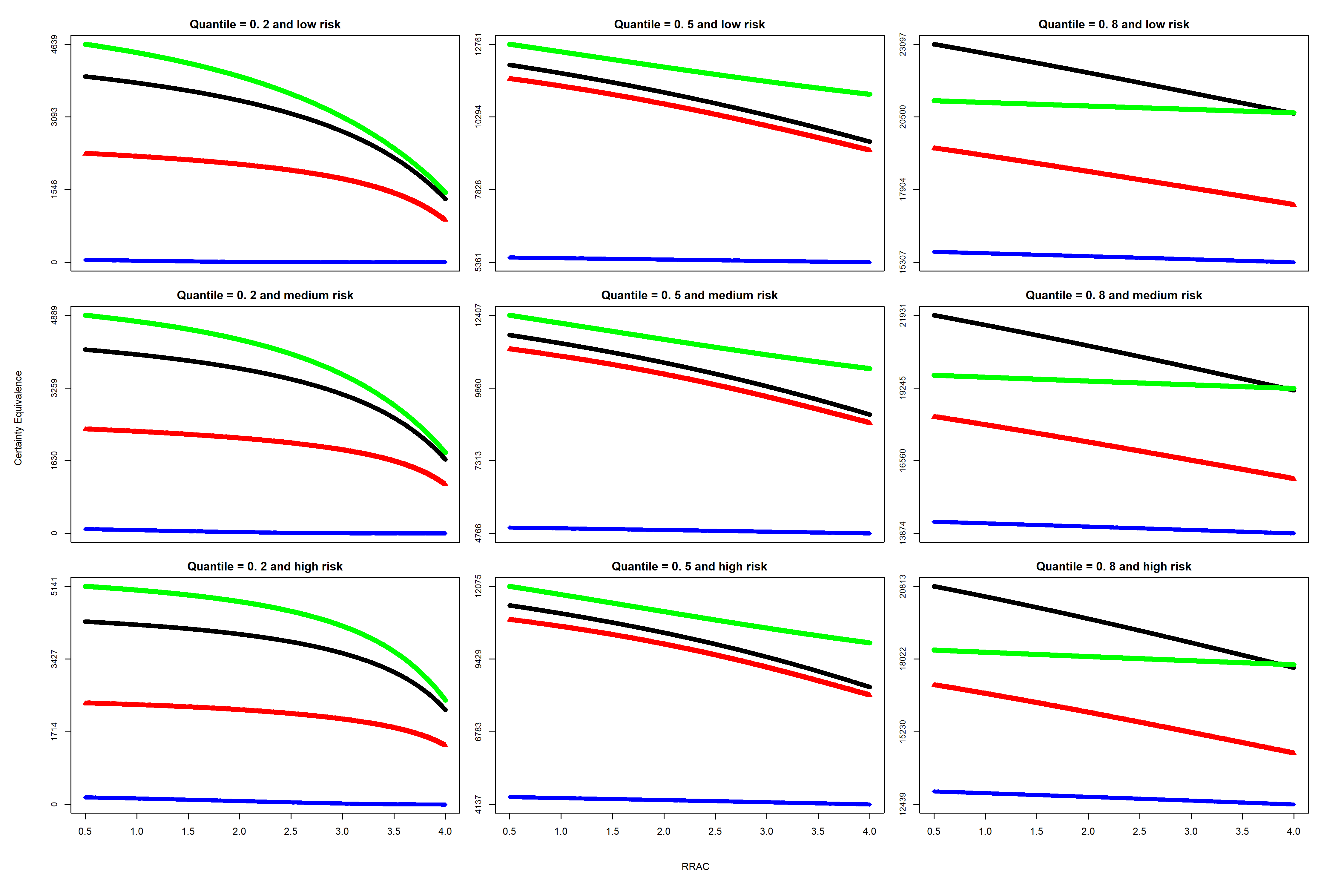}
\caption{Simulation with price 11.5 imposed.}
\label{fig:sen1}
\end{figure}


\begin{figure}[H]
\includegraphics[scale=.35]{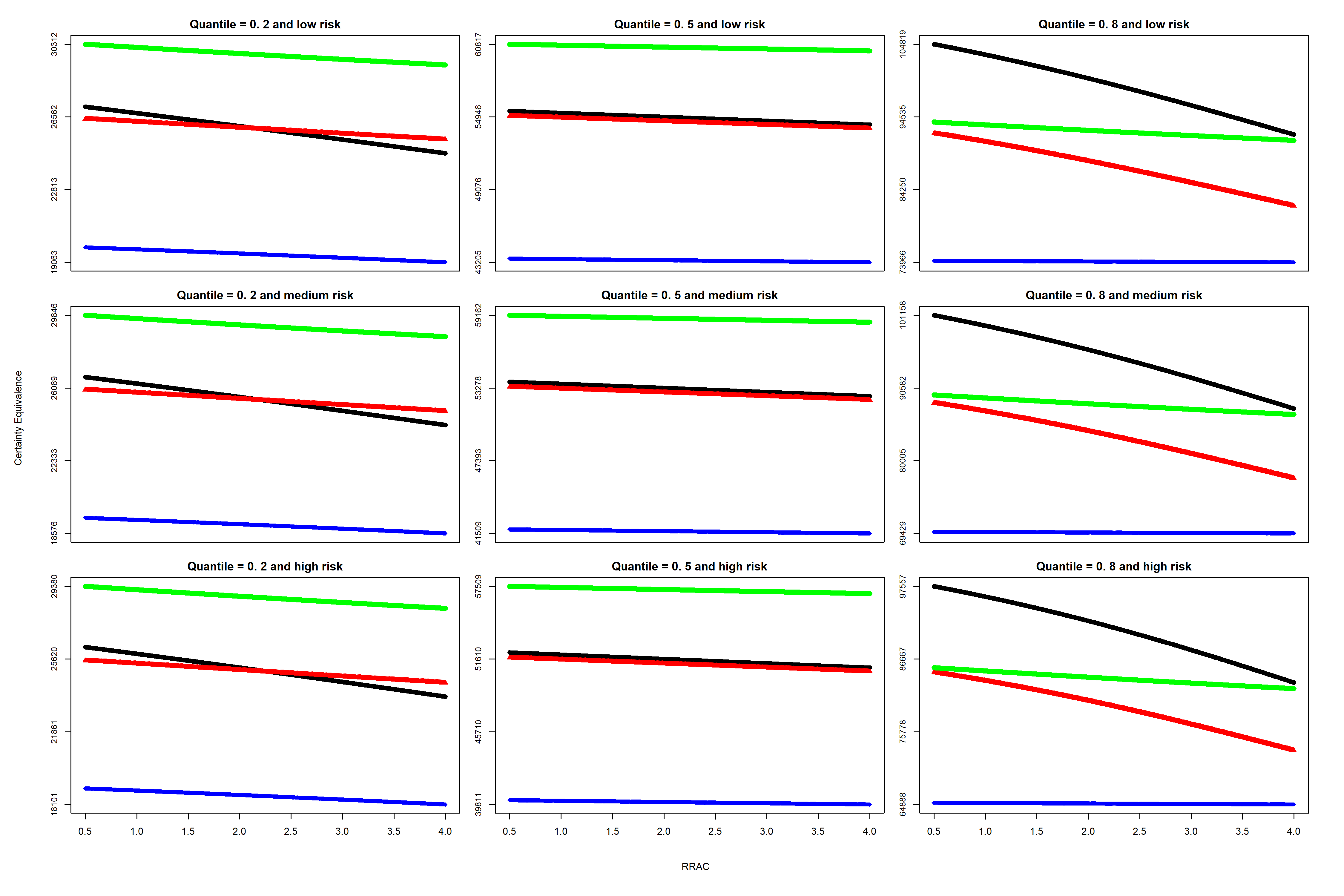}
\caption{Simulation with price 30 imposed.}
\label{fig:sen2}
\end{figure}


\end{document}